\documentstyle{article}

\begin{document}

\font\smc=cmcsc10
\hoffset-2cm

\centerline{\bf Quantum White Noise with Singular Non-Linear Interaction}
\bigskip
\centerline{L. Accardi\footnote{Graduate School of
Polymathematics, Nagoya University, Chikusa--ku, Nagoya, 464--01, Japan},
I.V. Volovich\footnote{Steklov Mathematical Institute of Russian
Academy of Sciences, Gubkin St.8, 117966 Moscow, Russia, 
email:volovich@arevol.mian.su}} \bigskip \centerline{Centro Vito Volterra, 
Universit\`a degli Studi di Roma ``Tor Vergata'' -- 00133, Italia    } 
%\centerline{$^{1}$ }
\bigskip\bigskip

\centerline{\bf Abstract}\bigskip\bigskip

 A model of a system driven by
 quantum white noise with
singular quadratic self--interaction is considered and
an exact solution
for the evolution operator  is found.
It is shown that the renormalized
square of the squeezed classical white noise is equivalent
to the quantum Poisson process.
 We describe how  equations
driven by nonlinear functionals of white noise can be derived
in  nonlinear quantum optics by using the stochastic
 approximation.
\bigskip

\bigskip
\vfill\eject

Quantum white noise has emerged in quantum optics [1,2] and it has been
widely studied in quantum theory [1,2], in infinite dimensional analysis [3]
and in quantum probability [4,5]. Ordinary white noise differential
equations describe quantum fluctuations in quantum optics, in laser
theory, in atomic physics, in the theory
of quantum measurement and in other topics and they are {\it linear}
with respect to white noise in the sense that the typical
equation for the evolution
operator $U_t$
has the form [1,2,4]
$${dU_t\over dt}\,=-i(F_tb_t^++F_t^+b_t)U_t\eqno(1)$$
Here $F_t$ is an operator describing the system (for example, atom)
and
$b_t$ and $b^+_t$ are quantum white noise operators,
$$[b_t,b^+_s]=\delta(t-s)\ ,\quad[b_t,b_s]=0$$

In this note we attempt to consider
white noise with {\it nonlinear} interaction. The motivation
for such a consideration is that if
fluctuations in a system are rather large
then they can produce a white noise
 with  nonlinear interaction. It is not clear apriori
what it means to have white noise with nonlinear interaction
because we want the evolution operator $U_t$ to be a bona fide unitary
operator and not a distribution.
Therefore we first consider a simple exactly solvable model
with such interaction and then discuss how one can derive
equations driven by nonlinear white noise in nonlinear quantum optics.

The first step to study
such noises is to investigate quantum white noise with
quadratic interaction. In this note we shall consider a model with the
following equation for the evolution operator $U_t$:
$${dU_t\over
dt}\,=-i[\omega_tb^+_tb_t+g_t(b^{+2}_t+b^2_t)+c_t]U_t\eqno(2)$$
where $\omega_t$,
     $g_t$ and $c_t$ are functions of time $t$. We shall
demonstrate that the singularity of the Hamiltonian imposes a
restriction to these functions which does not arise in the case
of regular Hamiltonians.

The consideration of models of quantum white noise with a non--linear
singular interaction like (2) was out of reach in the known approach [6]
to the quantum stochastic calculus.
It became possible only recently with the development of the
new white noise
approach to quantum stochastic calculus [4,5] in which methods of
renormalization theory have been used.
Actually already equation (1) requires a regularisation
as it will be discussed below.

As any model with quadratic interaction the model (2) is exactly
integrable in some sense. What makes the consideration of the model
interesting is the singular character of the interaction which involves
products of operators $b_t$ at the same time, like $b^2_t$, and it is
not clear a priori that such products have a meaning at all.
In fact the model (2) shows certain  surprising properties.
Even
after having given a meaning to equation (2) one cannot naively
interpret the formal Hamiltonian in (2) as a usual self--adjoint
operator. For example the model (2) with $\omega_t=0$ and $g_t=$
constant has been considered in [7] and it has been shown that, even
after renormalization the solution of (2) is not unitary. The results of
the present note show that for a special region of values of the
parameters $\omega_t$, $g_t$ and $c_t$ the unitarity of the solution can
be guaranteed and in fact one has an explicit and simple  form for it.
Unitarity of the solution is proved for all values of the parameters
below a certain threshold. Strangely enough this threshold
corresponds exactly to the square of {\it classical} white noise.
It seems interesting to have an exactly solvable dynamical model
driven by a non--linear white noise term because it can be instructive for
the consideration of more realistic models.

We solve equation  (2) by using a Bogoliubov transformation, that is we
look for a real function $\theta_t$ such that, defining the operators
$a_t$ and $a^+_t$ by
$$b_t=a_tch\theta_t-a^+_tsh\theta_t\ ,\quad b^+_t=a^+_tch\theta_t-
a_tsh\theta_t\eqno(3)$$
one has
$$\omega_tb^+_tb_t+g_t(b^{+2}_t+b^2_t)=
\Omega_ta^+_ta_t+\kappa_t\delta(0)\eqno(4)$$
for appropriate choice of $\Omega_t $ and $\kappa_t$.  In (4) a formal
infinity appears as the $\delta$--function in zero $\delta(0)$. We remove it
by the renormalization, that is we choose the function $c_t$ in (1)
as $c_t=-\kappa_t\delta(0)$ so that one has
$$\omega_tb^+_tb_t+g_t(b^{+2}_t+b^2_t)+c_t=\Omega_ta^+_ta_t$$
and the renormalization constant $c_t$ does not alter the dynamics.

Notice that (3) implies that
$$[a_t,a^+_s]=\delta(t-s)\ ,\quad[a_t,a_s]=0$$
so the operators $a_t,a^+_s$ define a {\it squeezed white noise}.
One easily proves that such a real function $\theta_t$ must satisfy
the equation
$${g_t\over\omega_t}=
{sh\theta_tch\theta_t\over sh^2\theta_t+ch^2\theta_t}\eqno(5)$$
Therefore the only restriciton on the real functions $g_t$ and $\omega_t$ is
that $|g_t/\omega_t|<1/2$. Under this condition one deduces the
expression of $\Omega_t$ and $\kappa_t$
$$\Omega_t={\omega_t\over sh^2\theta_t+
ch^2\theta_t}\eqno(6)$$
$$\kappa_t=-{\omega_tsh^2\theta_t\over sh^2\theta_t
+ch^2\theta_t}$$

Let $T$ be the formal unitary operator of the Bogoliubov transformation
$$T^+b_tT=a_t\ ,\quad T^+b^+_tT=a^+_t$$

The Bogoliubov transformation (3) is not unitary represented in the
original Fock space ${\cal H}_b$ for $b$--particles with vacuum
$\psi_b$. The operator $T^+$ acts actually from ${\cal H}_b$ to another
Fock space ${\cal H}_a$ for $a$--particles with the vacuum $\psi_a=
T^+\psi_b$ (see for example [8]). One has the relation
$$(T^+\psi_b,a_{\tau_1}\dots a_{\tau_m}U_ta^+_{t_1}\dots a^+_{t_n}
T^+\psi_b)=(\psi_a,a_{\tau_1\dots}a_
{\tau_m}U_ta^+_{t_1}\dots a^+_{t_n}\psi_a)
\eqno(7)$$
In this relation in the left hand side one has the inner product in the
Fock space of $b$--particles ${\cal H}_b$ with operators $a_t$, $a^+_t$
being expressed in terms of $b_t$ and $b^+_t$ by formulas (3) and with
$U_t$ satisfying equation  (1). In the left hand side of (7) there are
meaningless operators $T$ that requires a regularization.

However the right hand side of the relation (7) is well defined. In the
right hand side one has the inner product in the Fock space ${\cal H}_a$
of $a$--particles with $U_t$ satisfying the following equation
$${dU_t\over dt}\,=-i\Omega_ta^+_ta_tU_t\eqno(8)$$
So we reduce the solution of
equation  (1) to the solution of equation  (8) with
$\Omega_t$ given by (6). Eq. (8) defines the squeezed quantum Poisson
process.

Now let us solve equation  (8). It is not in the normal form
and we have to use a regularisation. We assume the following
convenient regularization
$$a^+_ta_tU_t=\lim_{\varepsilon\to 0}{1\over2}\,a^+_t(a_{t-\varepsilon}
+a_{t+\varepsilon})U_t\eqno(9)$$
From (8) one gets the following equation for the normal simbol $\tilde
U_t=\tilde U_t(\alpha^+,\alpha)$ of the operator $U_t$ (about normal
simbols see, for example [9])
$${d\tilde U_t\over dt}\,=-i\lim_{\varepsilon\to0}\Omega_t
\alpha^+_t\left(\alpha_t+{1\over2}\,\left({\delta\over\delta
\alpha^+_{t-
\varepsilon}}\,+{\delta\over\delta\alpha_{t+
\varepsilon}}\right)\right)\tilde
U_t$$
the solution of which is
$$\tilde U_t=\exp\left\{-i\int^t_0\sigma_\tau\alpha^+_\tau\alpha_\tau
d\tau\right\}\eqno(10)$$
where
$$\sigma_t={\Omega_t\over1+{i\over2}\,\Omega_t}\eqno(11)$$
The operator $U_t$ with the normal simbol (10), (11) is unitary if
$\Omega_t$ is a real function.
The normal simbol (10) corresponds to the following stochastic
differential equation in the sense of [6]
$$dU_t=-idN(\sigma_t)U_t\eqno(12)$$

The regularisation we have used is not unique. A more general
regularization is
$$ a_tU_t=\lim_{\varepsilon\to 0}[ca_{t-\varepsilon}+
(1-c)a_{t+\varepsilon}]U_t$$
where $c$ is an arbitrary complex constant.
Then instead of (11) one gets
$$\sigma_t={\Omega_t\over1+ic\Omega_t}$$
In this case the operator $U_t$ is unitary if
$c={1\over 2}+ix$ where $x$ is an arbitrary real number.

Finally by using (10)--(11) and performing the Gaussian funcitonal
integrals for the normal simbol one gets the following expression for
correlators
$$(\psi_a,\exp\left\{\int f_1(\tau)a_\tau d\tau\right\}U_t\exp
\left\{\int f_2(\tau)a^+_\tau d\tau\right\}\psi_a)=$$
$$=\exp\left\{\int f_1(\tau){1+{i\over2}\,\Omega_\tau\over1-{i\over
2}\,\Omega_\tau}\,f_2(\tau)d\tau\right\}\eqno(13)$$
where $f_1$ and $f_2$ are arbitrary functions.

Note that by using the similar regularisation for equation  (1)
$$U_t=1-i\lim_{\varepsilon\to 0}\int_0^t[F_{\tau}b_t^++F_{\tau}^+
(cb_{\tau-\varepsilon}+(1-c)b_{\tau+\varepsilon})]U_{\tau} d\tau
$$
one can write it in the normal form as
$${dU_t\over dt}\,=-i(F_tb_t^+U_t+F_t^+U_tb_t-icF^+_tF_tU_t)
\eqno(14)$$

Now let us comment about the unitarity condition in the model (2). The
operator $U_t$ (8) is unitary in the Hilbert space ${\cal H}_a$ for any
real function $\Omega_t$. However to come to this $U_t$ we have used the
Bogoliubov transformation with the restriction 
$|g_t/\omega_t|<1/2$ to
the parameters in the original model (2).
In the limit $g_t/\omega_t\to1/2$ there is no
dynamics because one gets $\Omega_t\to 0$ and $U_t\to 1$.
The critical case
$g_t/\omega_t =1/2$ corresponds exactly to the equation
$$dU_t/dt=-ig_t:(b^+_t+b_t)^2:U_t\eqno(15)$$
driven by the renormalized square of the {\it classical} white noise
$w_t=b_t^++b_t$.
So we get that eq. (15) leads only to the trivial
evolution operator $U_t=1$. To obtain a nontrivial evolution
one has to make a renormalization of the parameter $g_t$.
Let us consider the  following model
$${dU_t\over
dt}\,=-i{f_t\over\epsilon}[2b^+_tb_t+(1-{\epsilon^2\over2}
)(b^{+2}_t+b^2_t)+c_{\epsilon ,t}]U_t\eqno(16)$$ 
Eq. (16) is a regularisation of the square of the classical white noise
because in the formal limit $\epsilon \to0$
one gets eq.(15) with $g_t=f_t/\epsilon$.
In the limit $\epsilon\to0$ 
by using formulas (5) and (6) one gets $\Omega_t\to 2f_t$
and therefore the model (16)
is equivalent in this limit to the  model
describing the quantum Poisson process
$${dU_t\over dt}\,=-i2f_ta^+_ta_tU_t$$
So we have demonstrated that the model with the 
renormalized square of the squeezed classical white noise
has a meaning and it defines the quantum Poisson
process.

It would be interesting to study also the case when
$|g_t/\omega_t|>1/2$.
Here we would like to
mention a  possible relation of this question with a
non--associative Ito algebra of stochastic differentials introduced in
[4].
The linear stochastic differentials are defined as
$$ dB_t=b_tdt, \quad dB_t^+=b_t^+dt\eqno(17)$$
and they satisfy the quantum Ito multiplication rule
$dB_tdB_t^+=dt$ that can be derived from (17) by using the formal
identity
$$\delta (0)dt=1\eqno(18)$$
and also $b_t^+b_t(dt)^2=0$. Eq. (12) in these notations 
takes the form of the quantum stochastic differential equation [6]
$$dU_t\,=-i(F_tdB_t^+U_t+F_t^+U_tdB_t-icF^+_tF_tdtU_t)$$

To write equation  (2) as the quantum stochastic differential equation
we have to introduce  nonlinear stochastic differentials
$$dB_t^{(m,n)}=b_t^{+m}b_t^ndt$$
By using (18) and a renormalization prescription
one can get the following non-associative generalization
of the Ito multiplication rule
$$dB_t^{(m,n)}dB_t^{(k,l)}=nkdB_t^{(m+k-1,n+l1)}\eqno(19)$$
It is an important open problem to study the relation
of the algebra (19) with the unitarity of the evolution
operator in the model (2). 
 
Now let us discuss how quantum white noises with non--linear singular
interactions arise in the stochastic 
approximation to the usual Hamiltonian quantum
systems. Actually this is a rather general effect in theory of quantum
fluctuations [5]. 
In the stochastic limit of quantum theory white noise Hamiltonian
equations such as (1) are obtained as scaling limits of usual
Hamiltonian equations. 

One has to use the formalism of non-linear
quantum optics for the consideration of such problems
as how is a short pulse of squeezed light generated when
an intense laser pulse undergoes parametric downconversion in a 
traveling-wave configuration inside a 
non-linear crystal or how does such
a squeezed pulse undergo self-phase modulation as it propagates
in a non-linear optical fiber [1,10-13]. 
The Lagrangian describing the propagation of quantum light
through a nonlinear medium contains nonlinear terms in electric
field $E$ and magnetic field $B$ [10-13]:
$$L={1\over 2}(E^2-B^2)+\chi_{(2)}E^2 +\chi_{(3)}E^3+...$$
where $\chi_{(i)}$ are non--linear 
optical susceptibilities.
Such nonlinear terms lead to various non--linear quantum
noises that can be approximated by non--linear
quantum {\it white\/} noises. 

Let us consider a system interacting with quantum field
with the evolution equation of the form
$${dU(t)\over dt}\,=-i[\lambda^{2-n}\chi A(t)^n+...]U(t)\eqno(20)$$
 where
$$ A(t)=\int a(k)f(k)e^{i\omega (k)t}dk,\quad [a(k),a^+(p)]=\delta^3
(k-p)$$
$f(k)$ is a form--factor, $\chi$ is a constant and $\lambda$
is a small parameter.
Here the field operator $A(t)$ can be interpreted as a mode of
electromagnetic field in nonlinear quantum optics.

For example in the process of parametric down conversion
a photon of frequency $2\omega$ splits into two photons each
with frequency $\omega$ and in the simple model of parametric amplifier
where the pump mode at frequency $2\omega$ is classical
and the signal mode at frequency $\omega$ is described by the annihilation
operator $a$ one has the Hamiltonian of the form [1]
$$H=\omega a^+a-i\chi (a^2e^{2i\omega t}-a^{+2}e^{-2i\omega t})$$

In {\it the stochastic
(or $t/\lambda^2$--) limit} [14,4,5] one obtains from (14) an equation
of the type (2) with a singular interaction driven by a non-linear
quantum white noise. Indeed after the rescaling $t\to t/ \lambda^2$
 one gets equation
$${dU(t/\lambda^2)\over dt}\,=-i
[\chi\lambda^{-n} A({t\over \lambda^2})^n+...]U(t/\lambda^2)$$
which in the limit $\lambda \to 0$ becomes
$${dU_t\over dt}\,=-i[\chi b_t^n+...]U_t$$
because, as shown in [5],
$${1\over\lambda}A({t\over\lambda^2})\to b_t$$
in the sense that all the vacuum correlators of the left hand side
converge to the corresponding correlators of the right hand side.

Finally let us briefly comment about possible applications
of quantum white noise with nonlinear interaction
to non-linear quantum optics and to theory of quantum measurement.
The physical meaning of parameters $\omega_t$ and $g_t$
depends on the physical model. For instance
in the case of a short pulse propagating in a nonlinear
medium the parameters in the evolution equation
for the white noise are related with the 
nonlinear optical susceptibilities as it was discussed above.
If we take the sum of terms of the form (1) and (2) and $F_t$ are
function then such a model also is easily solved by means
of non-homogeneous Bogoliubov transformation. However
for a realistic model
not only $F_t$ should be operators but 
also the parameters $\omega_t$ and $g_t$ in principle
should be operators as well. The consideration of
such a model is very important but it is out of the scope of this note.

In the modern theory of quantum measurement [1] one
considers a system interacting with a linear
quantum white noise. For a system interacting
with a nonlinear medium one has to consider
quantum white noise with nonlinear interaction.
One expects that a dynamical model of
a system interacting with nonlinear white noise
can be interpreted as describing  the process
of self-measurement by analogy with the self-focusing
of a beam.

To conclude, the model (2) with quadratic singular interaction of
quantum white noise has been considered in this note.
We have demonstrated that the singular equation (2) leads to a well
defined unitary evolution operator (10), (11) and 
we have computed explicitly its matrix
elements (see (13)).
The model was solved under some restrictions to the parameters
$\omega_t$ and $g_t$ which deserve a further study. 
The results obtained in this exactly solved model could be useful for
investigation of more realistic and more complicated models with
quadratic as well as with higher order singular white noise interactions.

\bigskip

\centerline{Acknowledgments}\bigskip

One of the authors (I.V.) expresses his gratitude 
to the V.Volterra Center of the Roma University
Tor Vergata where this work was done for the hospitality.
He is also partially supported by RFFI-960100312\vfill\eject

\centerline{\smc Bibliography}\bigskip

\begin{description}
\item[1] D.F.Walls and G.J.Milburn, {\it Quantum Optics} (Springer, 1994).
\item[2] C.W. Gardiner, {\it Quantum Noise}, (Springer--Verlag, 1991).
\item[3] T. Hida, H--H. Kuo, J. Potthoff and L. Streit,
{\it White Noise: An Infinite Dimensional Calculus\/},
(Kluwer Academic, 1993).
\item[4] L. Accardi, Y.G. Lu and I. Volovich,
{\it Non--linear extensions of classical and quantum stochastic
calculus and essentially infinite dimensional analysis\/},
preprint Centro V. Volterra N. 268 (1996), Roma University
Tor Vergata, Rome, 1996. Proceedings of the Symposium
{\it Probability Towards Two Thousand\/}, Columbia University, New York,
2--6 October 1995, eds. L. Accardi, C. Heyde, (Springer, 1997).
\item[5] L. Accardi, Y.G. Lu and I. Volovich,
{\it Quantum Theory and Its Stochastic Limit\/}, (Oxford 
University Press, 1997).
\item[6] R.L. Hudson and K.R. Parthasarathy,
Comm. Math. Phys. 109, (1984) 301.
\item[7] L. Accardi, Y.G. Lu and N. Obata,
Towards a non-linear extension of stochastic calculus,
Preprint of Nagoya University, 1996.
\item[8] I.Segal, {\it Mathematical Problems of Relativistic Fields
}, (Providence: American Mathematical Society, 1963)
\item[9] F.A. Berezin, {\it The Method of Second Quantization},
(Academic Press, 1966).
\item[10] M. Hillery and E. Mlodinov, Phys. Rev. A30 (1984) 1860.
\item[11] P.D. Drummond, Phys. Rev. A 42 (1990) 6845.
\item[12] R.J. Glauber and M. Lowenstein, Phys. Rev. A 43 (1991) 467.
\item[13] I. Abram and E. Cohen, in: {\it Quantum Measurements
in Optics}, Eds. P.Tombesi and D.F.Walls, (Plenum Press, New York, 1992), 
p. 313.
\item[14] L. van Hove, Physica, 21, 517 (1955) 517.
\end{description}\end{document}